\documentclass{elsart}
\usepackage{epsfig}
\begin{document}
\runauthor{Cicero, Caesar and Vergil}
\begin{frontmatter}
\title{ Rapidly rotating axisymmetric neutron stars with quark cores } 
\vskip -0.0cm

\author[VECC]{Abhishek Mishra\thanksref{X}} 
\author[GDCT]{Partha Roy Chowdhury\thanksref{Y}} and
\author[VECC]{D.N. Basu\thanksref{Z}}
\vskip -0.0cm

\address[VECC]{Variable  Energy  Cyclotron  Centre, 1/AF Bidhan Nagar, Kolkata 700 064, India}
\address[GDCT]{Dept.of Physics, Govt.Degree College, Kamalpur, Dhalai, Tripura 799 285, India}

\thanks[X]{E-mail:abhishek.mishra@vecc.gov.in}
\thanks[Y]{E-mail:royc.partha@gmail.com}
\thanks[Z]{E-mail:dnb@vecc.gov.in}

\vskip -0.29cm
\begin{abstract}
\vskip -0.0cm

    We present a systematic study of the properties of pure hadronic and hybrid compact stars. The nuclear equation of state (EoS) for $\beta$-equilibrated neutron star matter was obtained using density dependent effective nucleon-nucleon interaction which satisfies the constraints from the observed flow data from heavy-ion collisions. The energy density of quark matter is lower than that of this nuclear EoS at higher densities implying the possibility of transition to quark matter inside the core. We solve the Einstein's equations for rotating stars using pure nuclear matter and quark core. The $\beta$- equilibrated neutron star matter with a thin crust is able to describe highly massive compact stars but find that the nuclear to quark matter deconfinement transition inside neutron stars causes reduction in their masses. Recent observations of the binary millisecond pulsar J1614-2230 by P. B. Demorest et al. \cite{De10} suggest that the masses lie within 1.97$\pm$0.04 M$_\odot$ where M$_\odot$ is the solar mass. In conformity with recent observations, pure nucleonic EoS determines that the maximum mass of NS rotating with frequency below r-mode instability is $\sim$1.95 M$_\odot$ with radius $\sim$10 kilometers. Although compact stars with quark cores rotating with Kepler's frequency have masses up to $\sim$2 M$_\odot$, but if the maximum frequency is limited by the r-mode instability, the maximum mass $\sim$1.7 M$_\odot$ turns out to be lower than the observed mass of 1.97$\pm$0.04 M$_\odot$, by far the highest yet measured with such certainty, implying exclusion of quark cores for such massive pulsars.
\end{abstract}
\vskip -0.0cm
\noindent
\vskip -0.0cm
\end{frontmatter}

\section{Introduction}

    The mass-radius relationship of neutron stars (NSs) is of prime importance to understand the high-density low-temperature regime of the hadronic equation of state (EoS). Depending on this relationship, certain models for the hadronic EoS can either be confirmed or ruled out. Several attempts have been made on measuring the radii and masses of NSs to constrain the uncertainties in the high density behavior of the EoS. The observations on double NSs \cite{Th99}, glitches in radio pulsars \cite{Li99}, thermal emission \cite{He09} from accreting NSs and from millisecond X-ray pulsars lead to constraints on mass-radius relationship of NSs. Recently the pressure of neutron star matter at supranuclear density is measured by \"Ozel et al. \cite{Oz08} directly from observations using advanced astrophysical techniques and NS atmosphere modeling. The pressure extracted from NS mass-radius data crucially constrains the extension of the EoS to high density low temperature regime for stellar matter ruling out those who fail to comply with the recent mass-radius data. The quark matter can support a star as massive as $\sim$2 M$_\odot$ only if the quarks are strongly interacting and are therefore not `free' quarks. To overcome this situation, Dexheimer et al. \cite{Dx10,Sc10} have recently employed a hadronic SU(3) sigma-omega model including Delta-resonances and hyperons to describe the properties of neutron star matter by softer equation of state. Delta-resonances have a repulsive vector potential which works to counteract gravity in a compact star. They successfully reproduce both the measured mass-radius relationship and the extrapolated EoS by slightly lowering the coupling strength of the Delta resonances to the vector mesons. 

    In our previous works, the density dependent M3Y effective interaction (DDM3Y) which provides a unified description of the elastic and inelastic scattering \cite{Gu05,Gu06}, cluster \cite{Ba03}, $\alpha$ \cite{Ba03,scb07,CSB06} and proton radioactivities \cite{BCS08}, the symmetric and asymmetric nuclear matter \cite{BCS08,CBS09,BCS09}, was employed to obtain nucleonic EoS of the $\beta$-equilibrated NS matter \cite{Ch10,PRC11}. At high densities, the energy density of this $\beta$-equilibrated charge neutral NS matter is higher than that of quark matter signalling the onset of deconfinement transition to quark matter inside the star. In the present work, we solve the Einstein's equations for rotating stars using pure nuclear matter without and with quark matter core. A systematic study of the static as well as rotating compact stars with quark matter inside is presented in view of the recent observations of the massive compact stars. We shall find later that the present EoS unlike other EoS \cite{La07,Gl08,La06,Sc06} can explain successfully the recently observed mass-radius data. The effect of the presence of the quark core on the determination of maximum mass of NS will be investigated for both static and rotating stars.

\section{Present observational status}

    With the energies and interaction rates foreseen at FAIR, the compressed baryonic matter (CBM) will create highest baryon densities in nucleus-nucleus collisions to explore the properties of superdense baryonic matter and the in-medium modifications of hadrons. The compact stars provide natural testing laboratory for highly compressed matter. The stiffness of the high-density matter controls the maximum mass of compact stars. The analyses of mass-radius data on NSs by \"Ozel et al. \cite{Oz08} favor smaller masses lying within 1.6-1.9 M$_\odot$ with radii 8-10 kilometers. Recent mass measurement of the binary millisecond pulsar J1614-2230 by P.B. Demorest et al. \cite{De10} rules out the EoS which fail to predict the masses within 1.97$\pm$0.04 M$_\odot$. Most of the currently proposed EoS \cite{La07,Gl08,La06,Sc06} involving exotic matter, such as kaon condensates or hyperons failed to produce such a massive star. The measured mass of PSR J1748-2021B, a millisecond pulsar in the Globular Cluster NGC 6440, is claimed to be as high as 2.74$^{+0.41}_{-0.51} $M$_\odot$ (2$\sigma$) \cite{Fr08}. Moreover, a pulsar rotating faster (e.g., PSR J17482446ad) than the limit set by the r-mode instability has already been observed \cite{He06}. Further observations and a better r-mode modeling may shed more light on this issue.

\section{Construction of compact star models}

    If rapidly rotating compact stars were nonaxisymmetric, they would emit gravitational waves in a very short time scale and settle down to axisymmetric configurations. Therefore, we need to solve for rotating and
axisymmetric configurations in the framework of general relativity. For the matter and the spacetime the following assumptions are made. The matter distribution and the spacetime are axisymmetric, the matter and the spacetime are in a stationary state, the matter has no meridional motions, the only motion of the matter is a circular one that is represented by the angular velocity, the angular velocity is constant as seen by a distant
observer at rest and the matter can be described as a perfect fluid. The energy-momentum tensor of a perfect fluid $T^{\mu\nu}$ is given by

\vspace{-0.0cm}
\begin{equation}
T^{\mu\nu} = (\varepsilon+P)u^\mu u^\nu-g^{\mu\nu}P
\label{seqn1}
\end{equation}
\noindent
where $\varepsilon$, $P$, $u^\mu$ and $g^{\mu\nu}$ are the energy density, pressure, four velocity and the metric tensor, respectively. To study the rotating stars the following metric is used

\vspace{-0.5cm}
\begin{eqnarray}
ds^2 = -e^{(\gamma+\rho)} dt^2 + e^{2\alpha} (dr^2+r^2d\theta^2) \nonumber\\
       + e^{(\gamma-\rho)} r^2 \sin^2\theta (d\phi-\omega dt)^2
\label{seqn2}
\end{eqnarray}
\noindent
where the gravitational potentials $\gamma$, $\rho$, $\alpha$ and $\omega$ are functions of polar coordinates $r$ and $\theta$ only. The Einstein's field equations for the three potentials $\gamma$, $\rho$ and $\alpha$ have been solved using the Green's-function technique \cite{Ko89} and the fourth potential
 $\omega$ has been determined from other potentials. All the physical quantities may then be determined from these potentials. Obviously, at the zero frequency limit corresponding to the static solutions of the Einstein's field equations for spheres of fluid, the present formalism yields the results for the solution of the Tolman-Oppenheimer-Volkoff (TOV) equation \cite{TOV39}. We use the `rns' code \cite{St95} for calculating the compact star properties which requires EoS in the form of energy density versus pressure along with corresponding enthalpy and baryon number density and since we are using various EoS for different regions, these are smoothly joined.  

\section{The crustal equations of state}

    The different regions of a compact star are governed by different EoS. These can be broadly divided into two regions: a crust that accounts for about 5$\%$ of mass and about 10$\%$ of the radius of a star and the core is responsible for the rest of the mass and radius of a star. The outer layers are a solid crust $\sim$ 1 km thick, consisting, except in the outer few meters, of a lattice of bare nuclei immersed in a degenerate electron gas. As one goes deeper into the crust, the nuclear species become, because of the rising electron Fermi energy, progressively more neutron rich, beginning (ideally) as $^{56}$Fe through $^{118}$Kr at mass density $\approx$4.3$\times$10$^{11}$ g cm$^{-3}$. At this density, the `neutron drip' point, the nuclei have become so neutron rich that with increasing density the continuum neutron states begin to be filled, and the lattice of neutron-rich nuclei becomes permeated by a sea of neutrons.

    The EoS that cover the crustal region of a compact star are Feynman-Metropolis-Teller (FMT) \cite{FMT49}, Baym-Pethick-Sutherland (BPS) \cite{BPS71} and Baym-Bethe-Pethick (BBP) \cite{BBP71}. The most energetically favorable nucleus at low densities is $^{56}$Fe, the endpoint of thermonuclear burning. The FMT is based on Fermi-Thomas model to derive the EoS of matter at high pressures and covers the outermost crust which is essentially made up of iron and a fraction of the electrons bound to the nuclei. The major difficulty in deriving the equation of state is the calculation of the electronic energy. At subnuclear densities, from about 10$^4$ g cm$^{-3}$ up to the neutron drip density 4.3$\times$10$^{11}$ g cm$^{-3}$ the EoS of BPS is applicable which includes the effects of the lattice Coulomb energy on the equilibrium nuclide. The domain from neutron drip density to about nuclear density 2.5$\times$10$^{14}$ g cm$^{-3}$, is composed of nuclei, electrons and free neutrons where EoS of BBP is applicable which is based on a compressible liquid drop model of nuclei with conditions that nuclei must be stable against $\beta$-decay and free neutron gas must be in equilibrium with neutrons in nuclei. 

\section{The $\beta$-equilibrated nuclear matter and quark matter EoS for compact star core}

    The nuclear matter EoS is calculated \cite{BCS08} using the isoscalar and the isovector components of M3Y interaction along with density dependence. The density dependence of the effective interaction, DDM3Y, is completely determined from nuclear matter calculations. The equilibrium density of the nuclear matter is determined by minimizing the energy per nucleon. The energy variation of the zero range potential is treated accurately by allowing it to vary freely with the kinetic energy part $\epsilon^{kin}$ of the energy per nucleon $\epsilon$ over the entire range of $\epsilon$. In a Fermi gas model of interacting neutrons and protons, the energy per nucleon for isospin asymmetric nuclear matter \cite{BCS08} is given by

\begin{equation}
 \epsilon(\rho,X) = [\frac{3\hbar^2k_F^2}{10m}] F(X) + (\frac{\rho J_v C}{2}) (1 - \beta\rho^{\frac{2}{3}})  
\label{seqn3}
\end{equation}
\noindent
where $k_F$=$(1.5\pi^2\rho)^{\frac{1}{3}}$ which equals Fermi momentum in case of symmetric nuclear matter (SNM), the isospin asymmetry $X=\frac{\rho_n-\rho_p}{\rho}$ with $\rho=\rho_n+\rho_p$, $\rho_n$ and $\rho_p$ being the neutron and proton number densities respectively, and the kinetic energy per nucleon $\epsilon^{kin}$=$[\frac{3\hbar^2k_F^2}{10m}] F(X)$ with $F(X)$=$[\frac{(1+X)^{5/3} + (1-X)^{5/3}}{2}]$. $J_v$=$J_{v00} + X^2 J_{v01}$, $J_{v00}$ and $J_{v01}$ represent the volume integrals of the isoscalar and the isovector parts of the M3Y interaction. The isoscalar $t_{00}^{M3Y}$ and the isovector $t_{01}^{M3Y}$ components of M3Y interaction potential are, respectively, given by $t_{00}^{M3Y}(s, \epsilon^{kin})=7999 \frac{\exp( - 4s)}{4s}-2134\frac{\exp( - 2.5s)}{2.5s}+ J_{00}(1-\alpha\epsilon^{kin})\delta(s)$ and $t_{01}^{M3Y}(s, \epsilon^{kin})=-4886\frac{\exp( - 4s)}{4s}+1176\frac{\exp( - 2.5s)}{2.5s}+J_{01}(1-\alpha\epsilon^{kin})\delta(s)$ where $J_{00}=-276$ MeVfm$^3$, $J_{01}=228$ MeVfm$^3$, $\alpha=0.005$ MeV$^{-1}$. The DDM3Y effective NN interaction is given by $v_{0i}(s,\rho, \epsilon) = t_{0i}^{M3Y}(s, \epsilon) g(\rho)$ where the density dependence $g(\rho) = C (1 - \beta \rho^{\frac{2}{3}})$ and the constants $C$ and $\beta$ of density dependence have been obtained from the saturation condition $\frac{\partial\epsilon}{\partial\rho} = 0$ at $\rho = \rho_{0}$ and $\epsilon = \epsilon_{0}$ where $\rho_{0}$ and $\epsilon_{0}$ are the saturation density and the saturation energy per nucleon respectively \cite{BCS08}. This EoS evaluated at the isospin asymmetry $X$ determined from the $\beta$-equilibrium proton fraction $x_\beta$ [$=\frac{\rho_p}{\rho}$], obtained by solving $\hbar c (3 \pi^2\rho x_\beta)^{1/3}= 4E_{sym}(\rho) (1 - 2 x_\beta)$, provides EoS for the $\beta$-equilibrated NS matter where nuclear symmetry energy $E_{sym}(\rho)$ is given by 

\begin{equation}
 E_{sym}(\rho)=\epsilon(\rho,1) -\epsilon(\rho,0).
\label{seqn4}
\end{equation}
\noindent
    
    The areas enclosed by the continuous and dashed lines in Fig.-1 correspond to the pressure regions for neutron matter consistent with the experimental flow data after inclusion of the pressures from asymmetry terms with weak (soft NM) and strong (stiff NM) density dependences, respectively \cite{Da02}. Although, the parameters of the density dependence of DDM3Y interaction are tuned to reproduce the saturation energy per nucleon $\epsilon_0$ and the saturation density $\rho_0$ of symmetric nuclear matter which are obtained from finite nuclei, the agreement of this EoS with the experimental flow data \cite{Da02}, where the high density behaviour looks phenomenologically confirmed, justifies its extrapolation to high density \cite{BCS08}. 

\begin{figure}[t]
\vspace{0.0cm}
\centerline{\epsfig{file=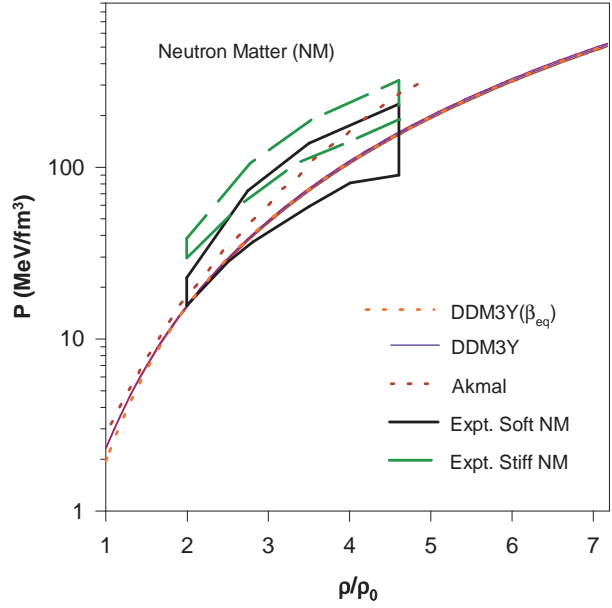,height=8.0cm,width=8.0cm}}
\caption
{ The pressures P of the $\beta$-equilibrated charge-neutral neutron star matter [DDM3Y($\beta_{\rm eq})$] and pure neutron matter for the present work [DDM3Y] and that of Akmal et al. \cite{Ak98} are plotted as functions of density $\rho/\rho_0$. The areas enclosed by the continuous and dashed lines correspond to the pressure regions for neutron matter consistent with the experimental flow data after inclusion of the pressures from asymmetry terms with weak (soft NM) and strong (stiff NM) density dependences, respectively \cite{Da02}.}
\label{fig1a}
\vspace {0.0cm}
\end{figure}

    For cold and dense quark (QCD) matter, the perturbative EoS \cite{Ku10} with two massless and one massive quark flavors and a running coupling constant, is used. The constant $B$ is treated as a free parameter, which allows to take into account non-perturbative effects not captured by the weak coupling expansion. In fact, using the free quark number density, one recovers the expression for the pressure in the original MIT bag model \cite{Ch74}, with $B$ taking the role of the bag constant. Due to physics criteria (e.g. requiring the energy density to be positive), the possible values for $B$ are, however, typically rather restricted, allowing to make quantitative statements that are not possible in the original MIT bag model.  

\begin{figure}[t]
\vspace{0.0cm}
\centerline{\epsfig{file=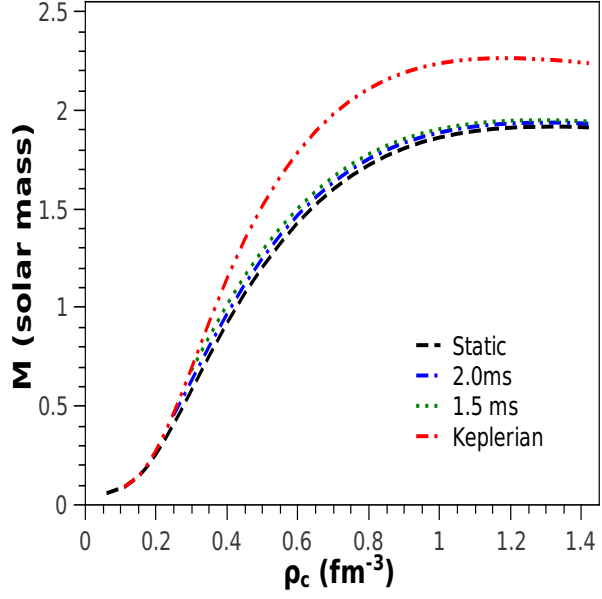,height=8.0cm,width=8.0cm}}
\caption
{ Variation of mass with central density for static and rotating neutron stars with pure nuclear matter inside.}
\label{fig1}
\vspace {0.0cm}
\end{figure}

\begin{figure}[t]
\vspace{0.0cm}
\centerline{\epsfig{file=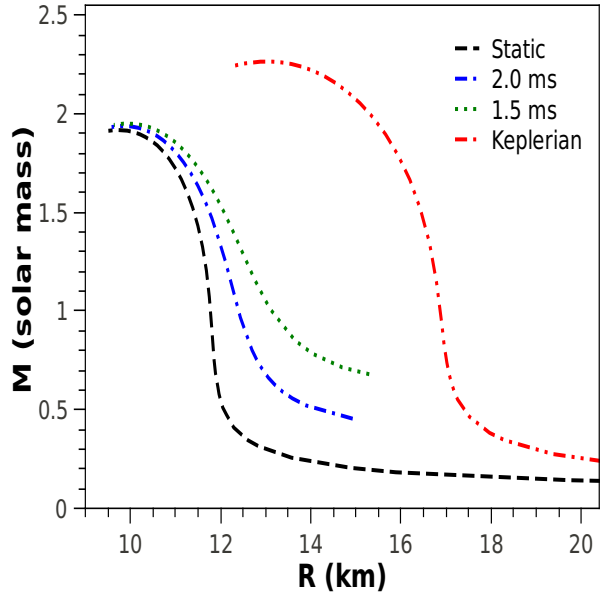,height=8.0cm,width=8.0cm}}
\caption
{ Mass-radius relationship for static and rotating neutron stars with pure nuclear matter inside.}
\label{fig2}
\vspace{0.0cm}
\end{figure}

\section{Calculations and Results}

    The rotating compact star calculations are performed using the crustal EoS, FMT + BPS + BBP upto number density of 0.0458 fm$^{-3}$ and $\beta$-equilibrated NS matter beyond. It is worthwhile to mention here that a star may not rotate as fast as Keplerian frequency due to r-mode instability. There have been suggestions that the r-mode instability may limit the time period to 1.5 ms \cite{St06}. However, a pulsar rotating faster (e.g., PSR J17482446ad) than this limit has already been observed \cite{He06}. The variation of mass with central density for static and rotating neutron stars at Keplerian limit and also maximum frequencies limited by the r-mode instability with pure nuclear matter inside is shown in Fig.-2. In Fig.-3, the mass-radius relationship for static and rotating neutron stars at Keplerian limit and also at maximum frequencies limited by the r-mode instability with pure nuclear matter inside is shown. Fig.-3 depicts that NSs with pure nuclear matter inside, the maximum mass for the static case is 1.92 M$_\odot$ with radius $\sim$9.7 km and for the star rotating with Kepler's frequency it is 2.27 M$_\odot$ with equatorial radius $\sim$13.1 km \cite{Ch10}. However, for stars rotating with maximum frequency limited by the r-mode instability, the maximum mass turns out to be 1.95 (1.94) M$_\odot$ corresponding to rotational period of 1.5 (2.0) ms with radius about 9.9 (9.8) kilometers \cite{PRC11}.
    
\begin{figure}[t]
\vspace{0.0cm}
\centerline{\epsfig{file=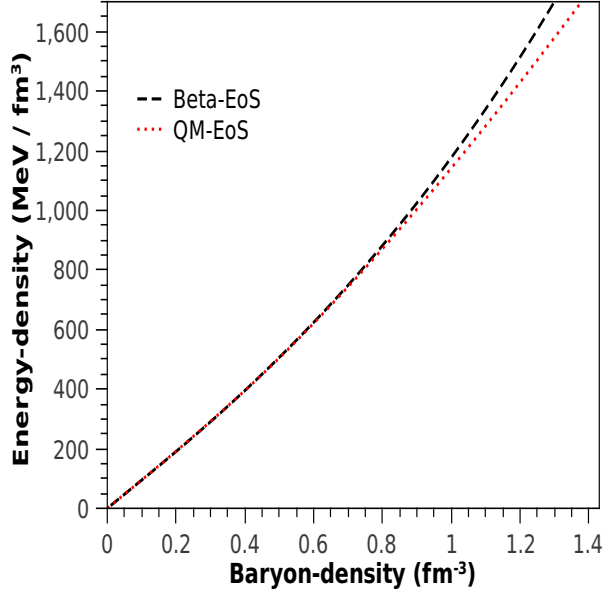,height=8.0cm,width=8.0cm}}
\caption
{ The EoS of the $\beta$-equilibrated charge neutral neutron star matter and the quark matter EoS.}
\label{fig3}
\vspace {0.0cm}
\end{figure}

    The energy density of the quark matter is lower than that of the present EoS for the $\beta$-equilibrated charge neutral NS matter at densities higher than 0.405 fm$^{-3}$ for bag constant $B^{\frac{1}{4}}$=110 MeV \cite{Ku10} implying presence of quark core. The energy densities of the present EoS for the $\beta$-equilibrated charge neutral NS matter and the quark matter EoS for bag constant $B^{\frac{1}{4}}$=110 MeV are shown in Fig.-4 as functions of baryonic densities. For lower values of bag constant such as $B^{\frac{1}{4}}$=89 MeV, energy density for our EoS is lower and makes a cross over with the quark matter EoS at very high density $\sim$1.2 fm$^{-3}$ causing too little quark core and therefore we choose $B^{\frac{1}{4}}$=110 MeV for representative calculations. The common tangent is drawn for the energy density versus density plots where pressure is the negative intercept of the tangent to energy density versus density plot. However, as obvious from Fig.-4, the phase co-existence region is negligibly small which is represented by part of the common tangent between the points of contact on the two plots \cite{Pe93} implying constant pressure throughout the phase transition. The variation of mass with central density for static and rotating neutron stars at Keplerian limit and also maximum frequencies limited by the r-mode instability with nuclear and quark matter inside is shown in Fig.-5. In Fig.-6, the mass-radius relationship for static and rotating neutron stars at Keplerian limit and also at maximum frequencies limited by the r-mode instability with nuclear and quark matter inside is shown. Fig.-6 depicts that when quark core is considered, the maximum mass for the static case is 1.68 M$_\odot$ with radius $\sim$10.4 km and for the star rotating with Kepler's frequency it is 2.02 M$_\odot$ with equatorial radius $\sim$14.3 km. In a similar study, it was concluded that compact stars with a quark matter core and an hadronic outer layer, can be as massive as 2.0 M$_\odot$ but stay below the pure quark stars and pure neutron stars \cite{We11}. However, they have used two different relativistic mean-field parameter sets TM1 and NL3 \cite{We11} to explore the influence of the hadronic part of the EoS whose high density behaviour do not satisfy the criteria extracted from the experimental flow data \cite{Da02}. For our case stars rotating with maximum frequency limited by the r-mode instability, the maximum mass turns out to be 1.72 (1.71) M$_\odot$ corresponding to rotational period of 1.5 (2.0) ms with radius about 10.7 (10.6) kilometers. 

\begin{figure}[t]
\vspace{0.0cm}
\centerline{\epsfig{file=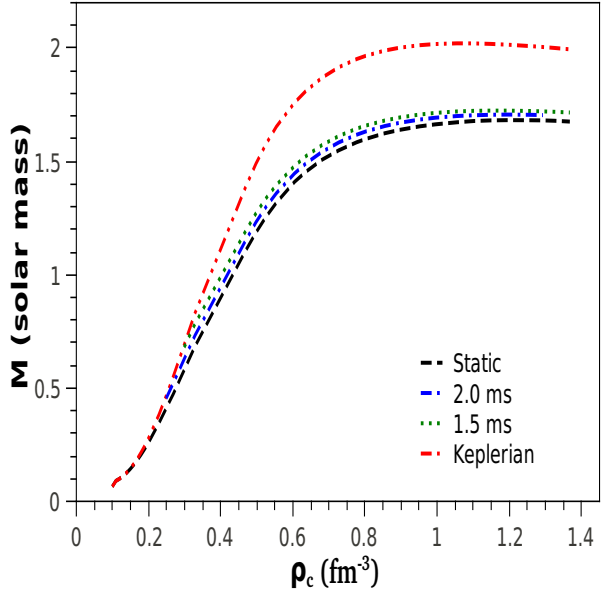,height=8.0cm,width=8.0cm}}
\caption
{ Variation of mass with central density for static and rotating neutron stars with nuclear and quark matter inside.}
\label{fig4}
\vspace {0.0cm}
\end{figure}

\begin{figure}[t]
\vspace{0.0cm}
\centerline{\epsfig{file=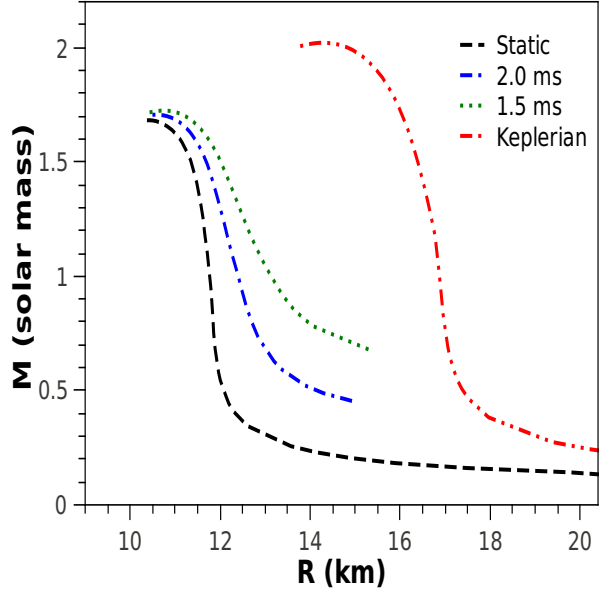,height=8.0cm,width=8.0cm}}
\caption
{ Mass-radius relationship for static and rotating neutron stars with nuclear and quark matter inside.}
\label{fig5}
\vspace{0.0cm}
\end{figure}

    It is worthwhile to mention here that the consideration of rotation at the Kepler frequency is not of much relevance for a comparison with observational data since even the shortest observed rotation period is still far from the Keplerian limit and does not result in a sufficient increase in the maximum mass. The recent observation of PSR J1614-2230 with a mass of 1.97$\pm$0.04 M$_\odot$ rotating with a period of 3.1 ms shows that either an EoS or transition to quark matter should be excluded that does not allow for star configurations with maximum mass reaching such high values. Since the hadronic EoS used in the present work is not soft (incompressibility 274.7$\pm$7.4 MeV) \cite{BCS08} at high densities and neutron star masses reach as high as $\sim$2 M$_\odot$ for such rotational frequencies, implies that theoretically robust quark matter EoS \cite{Ku10} used in this work is not stiff enough at high densities which was obtained from first principle calculations based on perturbation theory taking terms up to $O(\alpha_s^2)$ with quark chemical potentials and strange quark mass non-zero. The effects of quark pairing were incorporated by adding to the pressure a term accounting for the condensation energy of the cooper pairs in the color-flavor-locked
phase. However, a more detailed and realistic inclusion of the effect of pairing may make quark EoS stiffer. 

\section{Summary and Conclusion}

    In summary, the energy density of the present EoS for $\beta$-equilibrated charge neutral NS matter using DDM3Y effective NN interaction turns out to be higher than that of quark matter at densities above 0.405 fm$^{-3}$ implying possibility of quark core. We have applied our nucleonic EoS with a thin crust to solve the Einstein's field equations to determine the mass-radius relationship of neutron stars with and without quark cores. We have obtained the masses of neutron (hybrid) stars rotating with Keplerian frequencies, around 2.27 (2.02) M$_\odot$ with equatorial radii around 13 (14) kilometres. The result for NS without quark core is in excellent agreement with recent astrophysical observations. The neutron star matter can further undergo deconfinement transition to quark matter, thereby reducing compact star masses considerably. Although such hybrid compact stars rotating with Kepler's frequency have masses up to $\sim$2 M$_\odot$, but rotating with maximum frequency limited by the r-mode instability (or with 3.1 ms as observed for pulsar J1614-2230), the maximum mass $\sim$1.7 M$_\odot$ turns out to be lower than the observed mass of 1.97$\pm$0.04 M$_\odot$ and thus rules out quark cores for such massive pulsars but not for pulsars with masses $\sim$1.7 M$_\odot$ or less. Obviously, in order not to conflict with the mass measurement \cite{De10}, either there must be some mechanism to prevent nuclear matter to deconfine into quark matter or the quark EoS should be made stiffer by several possible realistic improvements \cite{Ku10}. The nucleon-nucleon effective interaction used in the present work, which is found to provide a unified description of elastic and inelastic scattering, various radioactivities and nuclear matter properties, also provides an excellent description of the $\beta$-equilibrated NS matter which is stiff enough at high densities to reconcile with the recent observations of the massive compact stars $\sim$2 M$_\odot$ while the corresponding symmetry energy is supersoft \cite{CBS09,BCS09} as preferred by the FOPI/GSI experimental data.

\end{document}